\documentclass[article]{IEEEtran}
\usepackage{amsfonts}
\usepackage{amsmath}
\usepackage{amsthm}
\usepackage{amssymb}
\usepackage{graphicx}
\usepackage[T1]{fontenc}
\usepackage{supertabular}
\usepackage{longtable}
\usepackage[usenames,dvipsnames]{color}
\usepackage{bbm}
\usepackage{fancyhdr}
\usepackage{breqn}
\usepackage{fixltx2e}
\usepackage{capt-of}
\usepackage{url}
\usepackage[framemethod=TikZ]{mdframed}
\usepackage{xcolor}
\usepackage{tikz}
\usepackage{endnotes}
\usepackage{supertabular}
\usepackage{tabularx}
\usepackage{lipsum}
\usepackage{soul}
\usepackage{colortbl}
\usepackage{supertabular}
\usepackage{longtable}
\usepackage{booktabs}

\newtheorem{remark}{Remark}

\newcommand{\mathsym}[1]{{}}
\newcommand{\unicode}[1]{{}}

\usepackage{grffile}
\usepackage[framemethod=TikZ]{mdframed}
\usepackage{framed}
\mdfsetup{%
	outerlinewidth=1,skipabove=20pt,backgroundcolor=yellow!20, outerlinecolor=none,
	innertopmargin=5pt,splittopskip=\topskip,skipbelow=\baselineskip, skipabove=\baselineskip,ntheorem,roundcorner=5pt}

\begin{document}

\title{\color{Brown} Branching epistemic uncertainty and thickness of tails }
\author{Nassim Nicholas Taleb\IEEEauthorrefmark{1}, Pasquale Cirillo\IEEEauthorrefmark{2}\\     
   \IEEEauthorblockA{  \IEEEauthorrefmark{1} NYU Tandon School \IEEEauthorrefmark{2} TU Delft  }
%\author{
   \thanks{\color{Brown}November 22, 2019. Corresponding author, NNT1@nyu.edu. A version of this article was presented by the corresponding author at Benoit Mandelbrot's Scientific Memorial at Yale University on April 30, 2011.    }
   }

%\author{Nassim Nicholas Taleb \\
% Pasquale Cirillo}
% \date{November 2019}

\maketitle
\thispagestyle{fancy}
\markboth{\textbf{FAT TAILS STATISTICAL PROJECT}}
\flushbottom % Makes all text pages the same height

\begin{mdframed}[backgroundcolor=yellow!10,	
nobreak=true,
%splitbottomskip=0.8cm,
%          splittopskip=0cm,
%          innerrightmargin=2cm,innertopmargin=1cm,%
          innerlinewidth=1pt,outerlinewidth=1pt,
          middlelinewidth=1pt,linecolor=brown, middlelinecolor=gray,
         % tikzsetting={draw=purple,line width=1pt,
               %        decorate,decoration={name=zigzag}},%
          ]

\begin{abstract}
This is an epistemological approach to errors in both inference and risk management, leading to necessary structural properties for the probability distribution.

Many mechanisms have been used to show the emergence of fat tails. Here we follow an alternative route, the epistemological one, using counterfactual analysis, and show how nested uncertainty, that is, errors on the error in estimation of parameters, lead to fattailedness of the distribution.

The results have relevant implications for forecasting, dealing with model risk and generally all statistical analyses. The more interesting results are as follows:
\begin{itemize}
\item The forecasting paradox: The future is fatter tailed than the past. Further, out of sample results should be fatter tailed than in-sample ones.
\item  Errors on errors can be explosive or implosive with different consequences. Infinite recursions can be easily dealt with, pending on the structure of the errors.
\end{itemize}
 We also present a method to perform counterfactual analysis without the explosion of branching counterfactuals.
\end{abstract}
\end{mdframed}

\section{The Regress Argument or Errors on Errors}
While there is a tradition of philosophical approaches to probability which includes most notably Laplace, Ramsey, Keynes, De Finetti, von Mises, Jeffreys, (see \cite{vonPlato} and \cite{Childers} for a review) and more recently Levi's magisterial \textit{Gambling with Truth} \cite{Levi}, such approaches and questions while influencing some branches of Bayesian inference, have not fully entered the field of quantitative risk management\footnote{See De Finetti\cite{deFinetti1, deFinetti2, deFinetti3} }. Yet epistemology, as a field, is a central element in statistical inference and risk management, %(Taleb and Pilpel, 2007) 
\cite{Taleb00, Taleb0, Rescher}. Fundamentally, philosophical inquiry is about handling such central questions about any inference as \textit{How do you know what you know?} \textit{ How certain are you about it?}, etc. This paper consider higher orders for such questions.

Now let's consider estimation. All estimates, whether statistical or obtained by some other method, are by definition imprecise (being "estimates"). An estimate must \textit{tautologically} have an error rate --otherwise it would not be an estimate, but a certainty or something linked to perfection. If we follow the logic, the \textit{error rate} itself is also an estimate --the methods used to estimate such error are themselves imprecise. There is no flawless way to estimate an error rate. And so forth.

By a regress argument, the inability to continuously re-apply this thinking about errors on errors fosters additional model uncertainty and unreliability not taken into account in the standard literature.\footnote{See \cite{Draper} for model risk.}

%On one hand the data we collect and the information we may use to elicit our beliefs are--most of the times--limited and not at all ``perfect", on the other each model, and each estimation technique, has limitations and a substantial degree of subjectivity \cite{deFinetti1, deFinetti2}, and this clearly affects the quality and the reliability of the results obtained. Model error falls within the larger and fundamental field of model risk \cite{Taleb2, Draper}, which is becoming a more and more relevant topic in statistical inference and modern risk management \cite{McNeil}.  
%
%From an epistemological point of view, first order errors, i.e. those on the estimates (of the parameters) of a model, can often just the tip of the iceberg. Not surprisingly, there is usually an even larger lack of understanding about second order errors--the errors on the errors--that is those related to the methods used to quantify first order errors. Paradoxically, even when they seem to care about first order errors, most researchers take for granted that their way of tackling the problem is error-free. A typical example is the use of higher moments to define confidence intervals for volatility under fat tails, when the existence of those moments is not guaranteed \cite{Embrechts,Taleb}. Clearly, not dealing correctly with second order errors increases model risk. 
%
%And what about the errors on the errors on the errors? 

While in practical applications, there is no problem with stopping the recursion in heuristically determined situations from past track records --where all errors have been determined through the test of time and survival and one has a satisfactory understanding of the structure and properties.  %, as we all agree that practice is different from pure speculation. 
However, refraining from caring about errors on errors should be explicitly declared as a subjective a priori decision that escapes quantitative and statistical methods. In other words we have to state and accept the subjectivity of the choice and the necessary effects on the overall model uncertainty. %As pointed out by de Finetti \cite{deFinetti3} when speaking about probability and statistics, and by innumerable philosophers like Bradley \cite{Bradley}, Nietzsche \cite{Nietzsche} and Russell \cite{Russell} to just name a few, objective truth is nothing but an invention. 
Using Isaac Israeli's words, as reported by Thomas Aquinas \cite{Aquinas}, \textit{``Veritas est adequatio intellectus et rei."}  

In what follows, we show how, taking the errors on errors argument to the limit, fat tails emerge from the layering of uncertainty. Starting from a completely non fat-tailed low-risk world, represented by the Normal distribution, we increase tail risk and generate fat tails by perturbing its standard deviation, introducing errors and doubts about its ``true" value. We show analytically how uncertainty induces fat tails, arguing that real life is actually even more extreme.\footnote{To use mathematical finance practitioners, "fat tails" in this context is any distribution with thicker tails than the Gaussian, not necessary a power-law. Hence this designation encompasses the subexponential and the power law classes, as well as any mixture of Gaussian with higher kurtosis than 3. }
\begin{mdframed}[backgroundcolor=yellow!10,	
nobreak=true,
%splitbottomskip=0.8cm,
%          splittopskip=0cm,
%          innerrightmargin=2cm,innertopmargin=1cm,%
          innerlinewidth=1pt,outerlinewidth=1pt,
          middlelinewidth=1pt,linecolor=brown, middlelinecolor=gray,
         % tikzsetting={draw=purple,line width=1pt,
               %        decorate,decoration={name=zigzag}},%
          ]

\begin{remark}
One of the contributions in this paper is the streamlining of counterfactuals.

Counterfactual analysis \cite{Lewis}, or the business school version of it, "scenario analysis", branch uncontrollably in an explosive manner, hampering projections many steps in the future --typically at a minimal rate $2^n$, where $n$ is the number of steps. We show that they can be structured analytically in a way that produces a single distribution of outcomes and allows variability. We manage to do do thanks to a rate of "error on error", which can be parametrized, hence allows for perturbations and sensitivity analyses.
\end{remark}

\end{mdframed}

\begin{mdframed}[backgroundcolor=yellow!10,	
nobreak=true,
%splitbottomskip=0.8cm,
%          splittopskip=0cm,
%          innerrightmargin=2cm,innertopmargin=1cm,%
          innerlinewidth=1pt,outerlinewidth=1pt,
          middlelinewidth=1pt,linecolor=brown, middlelinecolor=gray,
         % tikzsetting={draw=purple,line width=1pt,
               %        decorate,decoration={name=zigzag}},%
          ]
      \begin{remark}

          The mechanism by which uncertainty \textit{about probability} thickens the tails (by increasing the odds of tail events) is as follows.
          
          Assume someone tells you the probability of an event is \textit{exactly} zero. 
          
          You ask: "How do you know?"
          
          Answer: "I estimated it."
          
          Visibly if the person estimated it and there is, say, a 1\% error (symmetric), then probability has a lower bound of 1\%. Such uncertainty raises probability. It cannot be zero but some number higher than 1\%.
       	
      \end{remark}   
          
\end{mdframed}

\section{Layering Uncertainties} \label{layers}
Take a rather standard probability distribution, say the Normal. Assume that its dispersion parameter, the standard deviation $\sigma$, is to be estimated following some statistical procedure to get $\hat{\sigma}$. Such an estimate will nevertheless have a certain error, a rate of epistemic uncertainty, which can be expressed with another measure of dispersion: a dispersion on dispersion, paraphrasing the ``volatility on volatility" of option operators \cite{Derman, Dupire, Taleb}. This makes particularly sense in the real world, where the asymptotic assumptions \cite{Shao} usually made in mathematical statistics do not hold \cite{Taleb}, and where every model and estimation approach is subsumed under a subjective choice \cite{deFinetti3}.

Let $\phi(x;\mu,\sigma)$ be the probability density function (pdf) of a normally distributed random variable $X$ with known mean $\mu$ and unknown standard deviation $\sigma$. To account for the error in estimating $\sigma$, we can introduce a density $f_1(\hat{\sigma};\bar{\sigma},\sigma_1)$ over $\mathbb{R}^+$, where $\sigma_1$ represents the scale parameter of $\hat{\sigma}$ under $f_1$, and $\bar{\sigma_1}=\sigma$ its expected value. We are thus assuming that $\hat{\sigma}$ is an unbiased estimator of $\sigma$, but our treatment could also be adapted to the weaker case of consistency \cite{Shao}. In other words, the estimated volatility $\hat{\sigma}$ is the realization of a random quantity, representing the true value of $\sigma$ with an error term.

The unconditional law of $X$ is thus no longer that of a simple Normal distribution, but it corresponds to the integral of $\phi (x;\mu ,\sigma)$, with $\sigma$ replaced by $\hat{\sigma}$, across all possible values of $\hat{\sigma}$ according to $f_1(\sigma;\bar{\sigma}_1,\sigma_1)$. This known as a scale mixture of normals \cite{West}, and in symbols one has
\begin{mdframed}

\begin{equation}\label{eq1}
g_1(x)= \int _0^{\infty }\phi (x;\mu ,\hat{\sigma}) f_1(\hat{\sigma};\bar{\sigma}_1,\sigma_1) \mathrm{d}\hat{\sigma}.
\end{equation}
\end{mdframed}
Depending on the choice of $f_1$, that in Bayesian terms would define an a priori, $g_1(x)$ can take different functional forms.

Now, what if $\sigma_1$ itself is subject to errors? As observed before, there is no obligation to stop at Equation \eqref{eq1}: one can keep nesting uncertainties into higher orders, with the dispersion of the dispersion of the dispersion, and so forth. There is no reason to have certainty anywhere in the process. 

For $i=1,...,n$, set $\bar{\sigma}_i=E[\hat{\sigma}_{i}]$, with $\bar{\sigma}_1=\sigma$, and for each layer of uncertainty $i$ define a density $f_i(\hat{\sigma}_{i};\bar{\sigma}_{i},\sigma_i)$, with $\hat{\sigma}_1=\hat{\sigma}$. Generalizing to $n$ uncertainty layers, one then gets that the unconditional law of $X$ is now
\begin{mdframed}

\begin{eqnarray*}
g_n(x) &=& \int _0^{\infty }\int _0^{\infty }\cdots \int _0^{\infty }\phi (x;\mu ,\sigma)  f_1(\hat{\sigma}_1;\bar{\sigma}_1,\sigma_1)  \cdots \\
&&f_2(\hat{\sigma}_2;\bar{\sigma}_2,\sigma_2) \cdots  f_n(\hat{\sigma}_{n};\bar{\sigma}_{n_1},\sigma_n) \\
&&\mathrm{d}\hat{\sigma}_1 \, \mathrm{d}\hat{\sigma}_2 \cdots \mathrm{d}\hat{\sigma}_n.
\end{eqnarray*}
	
\end{mdframed}
This approach is clearly parameter-heavy and also computationally intensive, as it requires the specification of all the subordinated densities $f_i$ for the different uncertainty layers and the resolution of a possibly very complicated integral.

Let us consider a simpler version of the problem, by playing with a basic multiplicative process \`a la Gibrat \cite{Gibrat}, in which the estimated $\sigma$ is perturbed at each level of uncertainty $i$ by dichotomic alternatives: overestimation or underestimation. We take the probability of overestimation to be $p_i$, while that of underestimation is $q_i=1-p_i$.

Let us start from the true parameter $\sigma$, and let us assume that its estimate is equal to 
$$
\hat{\sigma}=\begin{cases}\sigma(1+ \epsilon_1) & \text{with }p_1 \\ 
\sigma(1- \epsilon_1) & \text{with }q_1 \end{cases},
$$
where $\epsilon_1\in [0,1)$ is an error rate (for example it could represent the proportional mean absolute deviation \cite{Taleb2}).

Equation \eqref{eq1} thus becomes
$$
g_1(x)= p_1 \phi (x;\mu ,\sigma(1+ \epsilon_1)) + q_1 \phi (x;\mu ,\sigma(1- \epsilon_1)).
$$

Now, just to simplify notation--but without any loss of generality--hypothesize that, for $i=1,...,n$, overestimation and underestimation are equally likely, i.e. $p_i=q_i=\frac{1}{2}$. Clearly one has that 
$$
g_1(x)= \frac{1}{2} \left( \phi (x;\mu ,\sigma(1+ \epsilon_1)) + \phi (x;\mu ,\sigma(1- \epsilon_1))\right).
$$

Assume now that the same type of uncertainty affects the error rate $\epsilon_1$, so that we can introduce $\epsilon_2\in [0,1)$ and define the element $(1\pm\epsilon_1)(1\pm\epsilon_2)$. Figure 1 gives a tree representation of the uncertainty over two (and possibly more) layers.

\begin{figure} \label{tree}
\includegraphics[width=\linewidth]{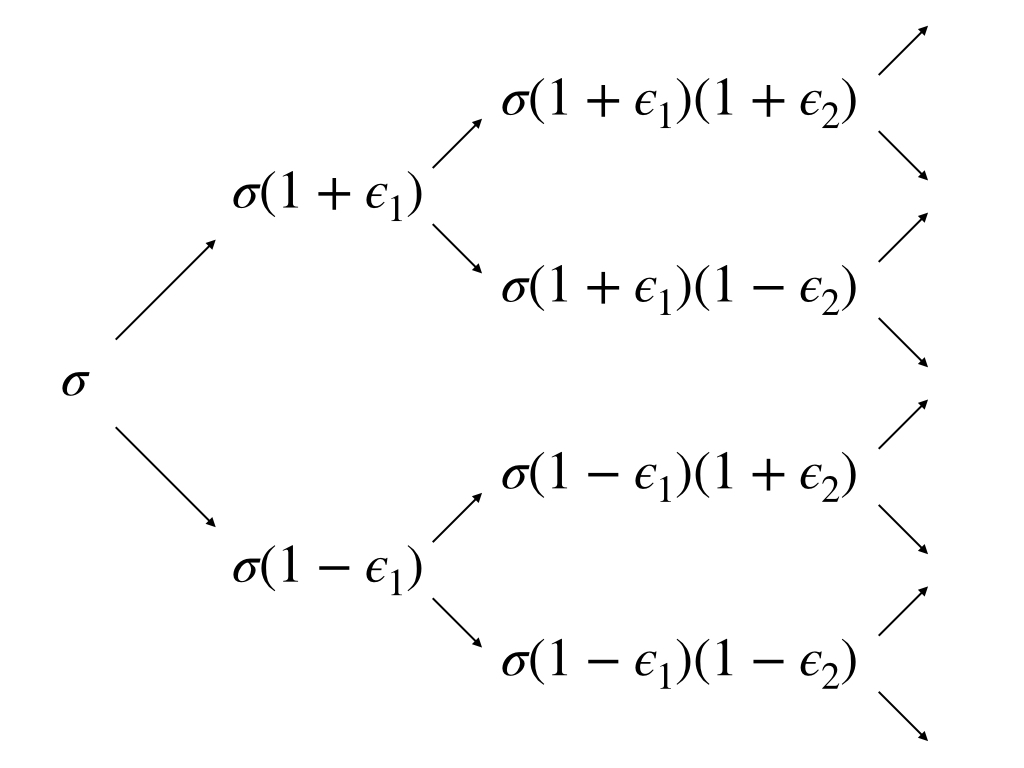}
\caption{Tree representation of the layers of uncertainty with dichotomic equiprobable states: overestimation and underestimation. One can see similarities with branching counterfactuals.}
\end{figure}

With two layers of uncertainty the law of $X$ thus becomes
\begin{eqnarray*}
g_2(x)&=&\frac{1}{4}\Bigg[\phi \bigg(x; \mu ,\sigma (1+\epsilon_1)(1+\epsilon_2)\bigg)\\
&+&\phi\bigg(x; \mu ,\sigma (1-\epsilon_1)(1+\epsilon_2)\bigg)\\
&+&\phi\bigg(x; \mu ,\sigma (1+\epsilon_1)(1-\epsilon_2)\bigg)\\
&+&\phi\bigg(x; \mu ,\sigma (1-\epsilon_1)(1-\epsilon_2)\bigg)\Bigg].
\end{eqnarray*}

While at the $n$-th layer, we recursively get

\begin{equation} \label{gn}
g_n(x)=2^{-n}\sum_{i=1}^{2^n} \phi\left(x;\mu,\sigma M_i^n\right),
\end{equation}
where $M_i^n$ is the $i$-th entry of the vector 
$$\textbf{M}^n=\left[\prod _{j=1}^n (1+\epsilon_j \mathbf{T}_{i,j})\right]_{i=1}^{2^n}
$$
with $\mathbf{T}_{i,j}$ being the $(i,j)$-th entry of the matrix of all the exhaustive combinations of $n$-tuples of the set $\{-1,1\}$, i.e. the sequences of length $n$ representing all the combinations of $1$ and $-1$. For example, for $n=3$, we have
\begin{equation*}
\mathbf{T}= \left[
\begin{array} {rrr}
 1 & 1 & 1 \\
 1 & 1 & -1 \\
 1 & -1 & 1 \\
 1 & -1 & -1 \\
 -1 & 1 & 1 \\
 -1 & 1 & -1 \\
 -1 & -1 & 1 \\
 -1 & -1 & -1
 \end{array} \right],
\end{equation*}

so that

%\begin{mdframed}
$$
	\mathbf{M}^3 =\left[
\begin{array}{c}
 (1-\epsilon_1) (1-\epsilon_2) (1-\epsilon_3) \\
 (1-\epsilon_1) (1-\epsilon_2) (1+\epsilon_3) \\
 (1-\epsilon_1) (1+\epsilon_2) (1-\epsilon_3) \\
 (1-\epsilon_1) (1+\epsilon_2) (1+\epsilon_3) \\
 (1+\epsilon_1) (1-\epsilon_2) (1-\epsilon_3) \\
 (1+\epsilon_1) (1-\epsilon_2) (1+\epsilon_3) \\
 (1+\epsilon_1) (1+\epsilon_2) (1-\epsilon_3) \\
 (1+\epsilon_1) (1+\epsilon_2) (1+\epsilon_3) \\
\end{array}
\right]
$$
%\end{mdframed}

and $M_1^3=\{(1-\epsilon_1) (1-\epsilon_2) (1-\epsilon_3)\}$.

Once again, it is important to stress that the various error rates $\epsilon_i$ are not sampling errors, but rather projections of error rates into the future. They are, to repeat, of epistemic nature.

Interestingly, Equation \eqref{gn} can be analyzed from different perspectives. In what follows we will discuss two relevant hypotheses regarding the error rates $\epsilon_i$, a limit argument based on the central limit theorem and an interesting approximation.

\section{Hypothesis 1: Constant error rate} \label{constant}

Assume that $\epsilon_1=\epsilon_2=...=\epsilon_n=\epsilon$, i.e. we have a constant error rate at each layer of uncertainty. What we can immediately observe is that matrix $\mathbf{M}$ collapses into a standard binomial tree for the dispersion at level $n$, so that

\begin{equation}
g_n(x)=2^{-n}\sum _{j=0}^n  \binom{n}{j}  \phi \left(x; \mu ,\sigma  (1+\epsilon)^j (1-\epsilon)^{n-j}\right).
\end{equation}

Because of the linearity of the sum, when $\epsilon$ is constant, we can use the binomial distribution to weight the moments of $X$, when taking $n$ layers of epistemic uncertainty. One can easily check that the first four raw moments read as

\begin{eqnarray*}
\mu^\prime_1&=&\mu, \\
\mu^\prime_2&=&\mu ^2+\sigma ^2 \left(1+\epsilon ^2\right)^n, \\
\mu^\prime_3&=&\mu ^3+3 \mu  \sigma ^2 \left(1+\epsilon ^2\right)^n, \\
\mu^\prime_4&=&3 \mu ^4+6 \mu ^2 \sigma ^2 \left(1+\epsilon ^2\right)^n+3 \sigma ^4 \left(1+6 \epsilon ^2+\epsilon ^4\right)^n. \\
\end{eqnarray*}

From these, one can then obtain the following notable moments \cite{Papoulis}:

\begin{eqnarray*}
\text{Mean}&:& \mu, \\
\text{Variance}&:&  \sigma ^2 \left(1+\epsilon ^2\right)^n, \\
\text{Skewness}&:& 0, \\
\text{Kurtosis}&:&3\left(\frac{\left(\epsilon ^4+6 \epsilon ^2+1\right)^n}{\left(1+\epsilon ^2\right)^{2 n} }-1\right). \\
\end{eqnarray*}

First notice that the mean of $X$ is both independent of $\epsilon$ and $n$: this is a clear consequence of the construction, for which $\mu$ is assumed to be known. For what concerns the variance, conversely, the higher the uncertainty ($n$ growing), the more dispersed is $X$; for $n\to \infty$, the variance explodes. Skewness, conversely, is not affected by uncertainty and the distribution of $X$ stays always symmetric. Finally kurtosis is a clear function of uncertainty, thus the distribution of $X$ becomes more and more leptokurtic as the layers of uncertainty increase, indicating a substantial thickening of the tails, hence a strong increase in risk.

Please observe that the explosion of the moments of order larger than one takes place for even very small values of $\epsilon\in [0,1)$, as $n$ grows to infinity. Even something as small as a $0.00001\%$ error rate will still lead to the invalidation of the use of \(\mathcal{L}^2\) distributions to study $X$. Once again, notwithstanding the error rate, the growth of uncertainty inflates tail risk, and such a behavior also occurs when $\epsilon_1\leq \epsilon_2\leq...\leq \epsilon_n$.

\begin{figure}[hb]
 \label{density}
	\includegraphics[width=\linewidth]{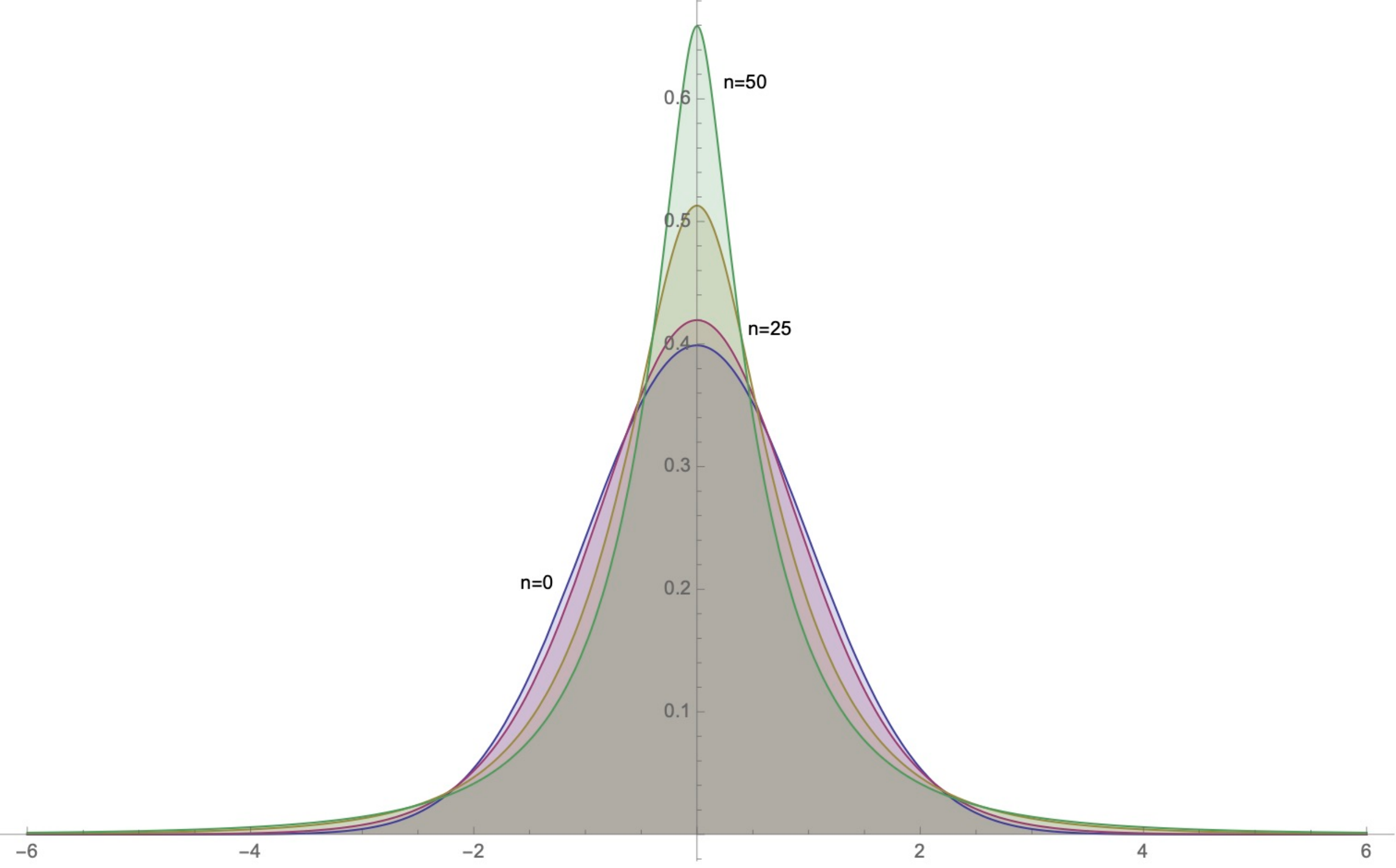}
\caption{Examples of the density of $X$, when $\mu=0$, $\sigma=1$ and $\epsilon=0.1$, for $n=0,5,10$. The larger $n$, the thicker the tail.}
\end{figure}

Figure 2 shows how the tails of $X$ get thicker as $n$ increases, compatibly with the explosion of moments. The larger $n$ the higher the kurtosis of $X$, so that its peak grows and so do the tails.

As observed before, in applications, there is no need to take $n$ large, it is totally understandable to put a cut-off somewhere for the layers of uncertainty, but such a decision should be taken a priori and motivated, in the philosophical sense.

Figure 3 shows the logplot of the density of $X$ when $\epsilon=0.1$, for different values of $n$. As expected, as $n$ grows, the tails of $X$ open up, tending towards a power law behavior, in a way similar to that of a risky lognormal with growing scaling parameter. Recall, however, that the first moment of $X$ will always be finite, suggesting that a pure power law behavior leading to an infinite-mean phenomenon will never take place. The result is also confirmed using other graphical tools (Zipf plot, Mean Excess Plot, etc.) like those discussed in \cite{Cirillo}.

\begin{figure}[hb]
 \label{loglog}
	\includegraphics[width=\linewidth]{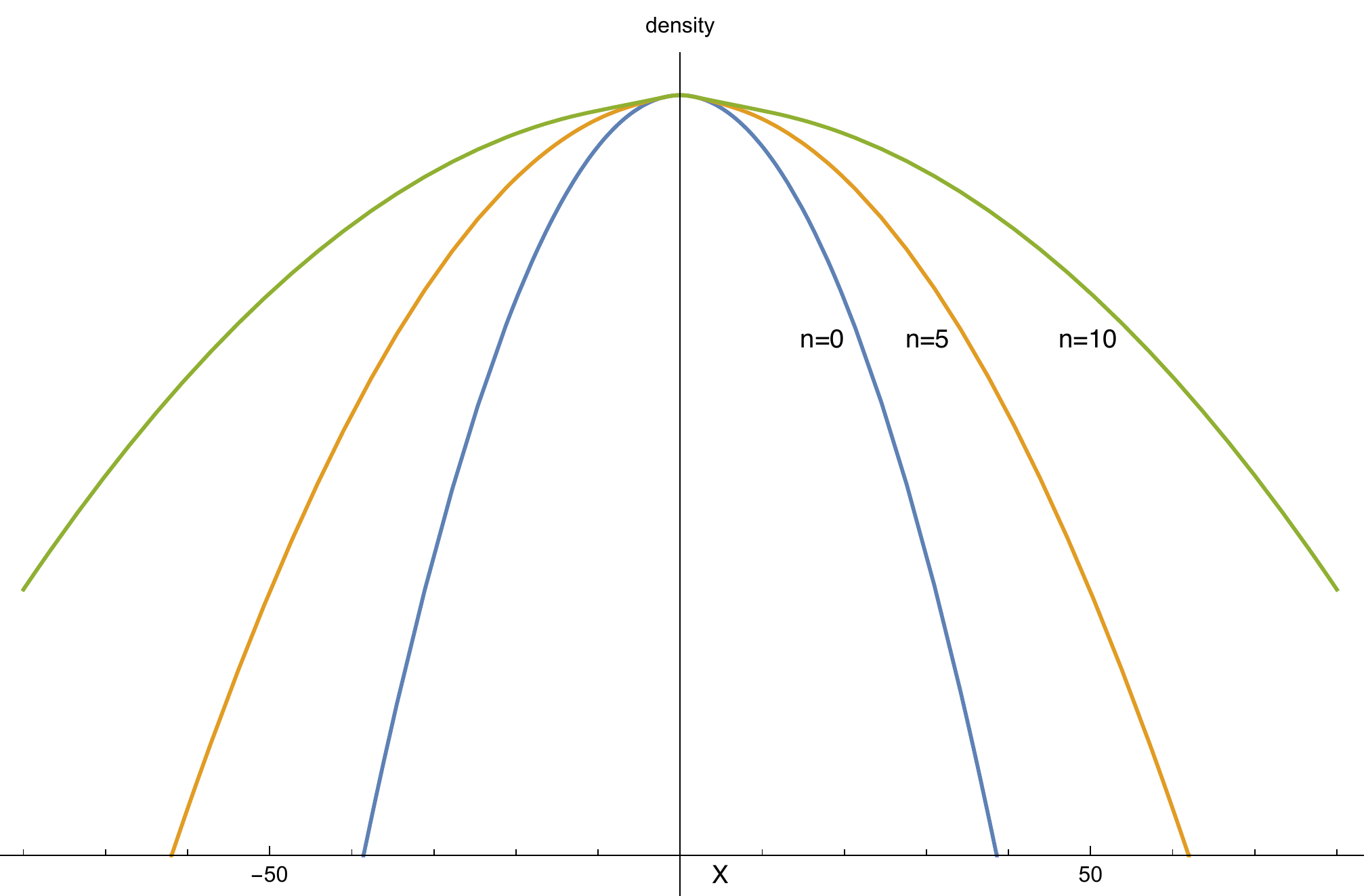}
\caption{Logplot of the density of $X$ for $\epsilon=0.1$ and different values of $n$. The larger $n$ the more the density tends to behaves like a power law in both tails, as suggested by the flattening.}
\end{figure}

It is then interesting to measure the effect of $n$ on the thickness of the tails of $X$. The obvious effect, as per Figure 3, is the rise of tail risk. 

Fix $n$ and consider the exceedance probability of $X$ over a given threshold $K$, i.e. the tail of $X$, when $\epsilon$ is constant. One clearly has

\begin{equation}
P(X\geq K)= \sum _{j=0}^n 2^{-n-1} \binom{n}{j} \text{erfc}\left(\frac{K}{\sqrt{2} \sigma  (1+\epsilon)^j (1-\epsilon)^{n-j}}\right), 
\end{equation}
where $\text{erfc}(z)=1-\frac{2}{\sqrt{\pi }}\int_0^ze^{-t^2}\mathrm{d}t$ is the complementary error function.

Tables \ref{T1} and \ref{T2} show the ratio of the exceedance probability of $X$ for different values of $K$ and $n$ over the benchmark represented by a simple normal with mean $\mu$ and variance $\sigma^2$, i.e. our starting point in case of no uncertainty on $\sigma$ (or, in other words, for $n=0$). The two tables differ for the value of $\epsilon$, equal to 0.1 in former, and 0.01 in latter. It is pretty clear how the layering of uncertainty, as $n$ grows, make the same tail probabilities grow dramatically. For example, the probability $P(X\geq10)$ is $3.62\times 10^{18}$ times larger than the corresponding probability for a $\text{Normal}(\mu,\sigma^2)$, when $n=25$ and $\epsilon$ is just $0.01$.

\begin{table}[hb]
\newcolumntype{a}{>{\columncolor{cyan!20}}c}
\begin{center}
\begin{tabular}{|a|ccc|}
\hline
\rowcolor{cyan!20} &  $K=3$ & $K=5$ &  $K=10$ \\
 \hline
 n=5 & 1.01724 & 1.155 & 7 \\
 n=10 & 1.0345 & 1.326 & 45 \\
 n=15 & 1.05178 & 1.514 & 221 \\
 n=20 & 1.06908 & 1.720 & 922 \\
 n=25 & 1.0864 & 1.943 & 3347 \\
 \hline
\end{tabular}
\end{center}
\caption{Ratio of the exceedance probability of $X$, when $\epsilon=0.1$, for different values of $K$ and $n$, over that of a normal random variable with mean $\mu$ and variance $\sigma^2$.}
 \label{T1}
\end{table}

\begin{table}[h]
\newcolumntype{a}{>{\columncolor{cyan!20}}c}
\begin{center}
\begin{tabular}{|a|ccc|}
\hline
 \rowcolor{cyan!20} &  $K=3$ & $K=5$ &  $K=10$ \\
 \hline
 n=5 & 2.74 & 146 & $1.09\times 10^{12}$ \\
 n=10 & 4.43 & 805 & $8.99\times 10^{15}$ \\
 n=15 & 5.98 & 1980 & $2.21\times 10^{17}$ \\
 n=20 & 7.38 & 3529 & $1.20\times 10^{18}$\\
 n=25 & 8.64 & 5321 & $3.62\times 10^{18}$ \\
 \hline
\end{tabular}
\end{center}
\caption{Ratio of the exceedance probability of $X$, when $\epsilon=0.01$, for different values of $K$ and $n$, over that of a normal random variable with mean $\mu$ and variance $\sigma^2$.}
 \label{T2}
\end{table}

\section{Hypothesis 2: Decaying error rates}

As oigurbserved before, one may have (actually one needs to have) a priori reasons to stop the regress argument and take $n$ to be finite. For example one could assume that the error rates vanish as the number of layers increases, so that $\epsilon_i\geq  \epsilon_j$ for $i<j$, and $\epsilon_i$ tends to 0 when $i$ approaches a given $n$. In this case, one can show that the higher moments tend to be capped, and the tail of $X$ less extreme, yet riskier than what one could naively think.

Take a value $\kappa \in [0,1]$ and fix $\epsilon_1$. Then, for $i=2,...,n$, hypothesize that $\epsilon_i=\kappa \epsilon_{i-1}$, so that $\epsilon_n=\kappa^n \epsilon_1$. For what concerns $X$, without loss of generality, set $\mu=0$.
With $n=2$, the variance of $X$ becomes
$$
\sigma^2 \left(1+\epsilon_1^2\right) \left(1+\kappa^2\epsilon_1^2 \right).
$$
For $n=3$ we get
$$
\sigma^2 \left(1+\epsilon_1^2\right) \left(1+\kappa^2\epsilon_1^2 \right) \left(1+\kappa^4\epsilon_1^2 \right).
$$
For a generic $n$ the variance is
\begin{equation}\label{var2} 
\sigma^2  \prod_{i=0}^{n-1}\left(1+\epsilon_1^2 \kappa^{2i} \right)=\sigma^2 \left[ -\epsilon^2;k^2\right]_n,
\end{equation}
where $[ a;q]_n= \prod_{i=0}^{n-1}\left(1+a q^i\right)$ is the $q$-Pochhammer function.

Going on computing moments, for the fourth central moment of $X$, one gets for example
$$
3\sigma^4\prod_{i=0}^{n-1}\left(1+6\epsilon^2\kappa^{2i}+\epsilon^4\kappa^{4i} \right).
$$
For $\kappa=0.9$ and $\epsilon_1=0.2$, we get a variance of $1.23\sigma^2$, with a significant yet relatively benign convexity bias. And the limiting fourth central moment is $9.88\sigma^4$, more than 3 times that of a simple Normal, which is $3\sigma^4$. Such a number, even if finite--hence the corresponding scenario is not as extreme as before--definitely suggests a tail risk not to be ignored.

For values of $\kappa$ in the vicinity of 1 and $\epsilon \downarrow 0$, the fourth moment of $X$ converges towards that of a Normal, closing the tails, as expected.

\section{A central limit theorem argument}

We now discuss a central limit theorem argument for epistemic uncertainty as a generator of thicker tails and risk. For doing so, we introduce a more convenient representation of the normal distribution, which will also prove useful in Section \ref{approximation}.

Consider again the real-valued normal random variable $X$, with mean $\mu$ and standard deviation $\sigma$. Its density function is thus
\begin{equation} \label{norm}
\phi(x; \mu,\sigma)=\frac{e^{-\frac{(x-\mu )^2}{2 \sigma ^2}}}{\sqrt{2 \pi } \sigma }.
\end{equation}
Without any loss of generality, let us set $\mu=0$. Moreover let us re-parametrize Equation \eqref{norm} in terms of a new parameter $\lambda = \frac{1}{\sigma^2}$, commonly called ``precision" in Bayesian statistics \cite{Bernardo}. The precision of a random variable $X$ is nothing more than the reciprocal of its variance, and, as such, it is just another way of looking at variability (actually Gauss \cite{Gauss} originally defined the Normal distribution in terms of precision). From now on, we will therefore assume that $X$ has density
\begin{equation} \label{precnorm}
\phi(x; \lambda)=\frac{\sqrt{\lambda } e^{-\frac{1}{2} \left(\lambda  x^2\right)}}{\sqrt{2 \pi }}.
\end{equation}

Imagine now that we are provided with an estimate of $\lambda$, i.e. $\hat{\lambda}$, and take $\hat{\lambda}$ to be close enough to the true value of the precision parameter. Assuming that $\lambda$ and $\hat{\lambda}$ are actually close is not necessary for our derivation, but we want to be optimistic by considering a situation in which who estimates $\hat{\lambda}$ knows what she is doing, using an appropriate method, checking statistical significance, etc.

We can thus write 
\begin{equation} \label{error1}
\lambda= \hat{\lambda} (1+\epsilon_1),
\end{equation}
where $\epsilon_1$ is now a first-order random error term such that $E[\epsilon_1]=0$ and $\sigma^2(\epsilon_1)<\infty$. Apart from these assumptions on the first two moments, no other requirement is put on the probabilistic law of $\epsilon_1$.

Now, imagine that a second order error term $\epsilon_2$ is defined on $1+\epsilon_1$, and again assume that it has zero mean and finite variance. The term $\epsilon_2$ may, as before, represent uncertainty about the way in which the quantity $1+\epsilon_1$ was obtained. Equation \eqref{error1} can thus be re-written as
\begin{equation} \label{error2}
\lambda= \hat{\lambda} (1+\epsilon_1)(1+\epsilon_2).
\end{equation}
Iterating the error on error reasoning we can introduce a sequence $\left\{\epsilon_i \right\}_{i=1}^n$ such that $E[\epsilon_i]=0$ and $\sigma^2(\epsilon_i) \in [c,\infty)$, $c>0$, so that we can write
\begin{equation} \label{errorn}
\lambda= \hat{\lambda} (1+\epsilon_1)(1+\epsilon_2)(1+\epsilon_3)\cdots(1+\epsilon_n).
\end{equation}
For $n\to \infty$, Equation \eqref{errorn} represents our knowledge about the parameter $\lambda$, once we start from the estimate $\hat{\lambda}$ and we allow for epistemic uncertainty, in the form of multiplicative errors on errors. The lower value $c>0$ for the variances of the error terms is meant to guarantee a minimum level of epistemic uncertainty at every level, and to simplify the application of the central limit argument below.

Now take the logs on both sides of Equation \eqref{errorn} to obtain
\begin{equation} \label{lerrorn}
\log(\lambda)= \log(\hat{\lambda})+ \log(1+\epsilon_1)+\log(1+\epsilon_2)+\cdots+\log(1+\epsilon_n).
\end{equation}
If we assume that, for every $i=1,...,n$, $|\epsilon_i|$ is small with respect to 1, we can introduce the approximation $\log(1+\epsilon_i)\approx \epsilon_i$, and Equation  \eqref{errorn} becomes
\begin{equation} \label{lerrornapprox}
\log(\lambda) \approx \log(\hat{\lambda})+ \epsilon_1+\epsilon_2+\cdots+\epsilon_n.
\end{equation}
To simplify treatment, let us assume that the error terms $\left\{\epsilon_i \right\}_{i=1}^n$ are independent from each other\footnote{In case of dependence, we can refer to one of the generalizations of the CLT \cite{Embrechts, Feller}.}. For $n$ large, a straightforward application of the Central Limit Theorem (CLT) of Laplace-Liapounoff \cite{Feller} tells us that $\log(\lambda)$ is approximately distributed as a $\text{Normal}(0,S^2_n)$, where $S^2_n=\sum_{i=1}^2 \sigma^2(\epsilon_i)$. This clearly implies that $\lambda \sim \text{Lognormal}(0,S^2_n)$, for $n \to \infty$. Notice that, for $n$ large enough, we could also assume $\hat{\lambda}$ to be a random variable (with finite mean and variance), but still the limiting distribution of $\lambda$ would be a Lognormal. For the reader interested in industrial dynamics, the above derivation should recall the so-called Gibrat law of proportionate effects for the modeling of firms' size \cite{Gibrat,Kleiber}.

From now on we drop the $n$ index from $S^2_n$, using $S^2$ and assuming that $n$ is large enough for the CLT to hold.

Epistemic doubt has thus a very relevant consequence from a statistical point of view. Using Bayesian terminology, the different layers of uncertainty represented by the sequence of random errors $\left\{\epsilon_i \right\}_{i=1}^n$ correspond to eliciting a Lognormal prior distribution on the precision parameter $\lambda$ of the initial Normal distribution. This means that, in case of epistemic uncertainty, the actual marginal distribution of the random variable $X$ is no longer a simple Normal, but a Compound Normal-Lognormal distribution, which we can represent as
\begin{equation} \label{compound}
g(x; \lambda,S)=\int_{0}^\infty f(x; \lambda) h(\lambda;S)\text{d} \lambda = \int_{0}^\infty \frac{e^{-\frac{\log ^2(\lambda )+\lambda  S^2 x^2}{2 S^2}}}{2 \pi  \sqrt{\lambda } S} \text{d} \lambda,
\end{equation}
where $h(\lambda;S)$ is the density of a $\text{Lognormal}(0,S^2)$ for the now random precision parameter $\lambda$. 

Notice that, for the properties of the Lognormal distribution \cite{Kleiber}, also the distribution of $\frac{1}{\sigma^2}$ is Lognormal. However, the use of the parametrization based on the precision of $X$ is convenient in view of the next section. 

In fact, despite its apparent simplicity, the integral in Equation \eqref{compound} cannot be solved analytically. This means that we are not able to obtain a closed form for the Compound Normal-Lognormal (CNL) distribution represented by $g(x; \lambda,S)$, even if its first moments can be obtained explicitly. For example the mean is equal to $0$ ($\mu$ in general), while the kurtosis is $3\left(e^{S^2}-1\right)$.

\section{The analytical approximation} \label{approximation}
The impossibility of solving Equation \eqref{compound} can be somehow by-passed by introducing an approximation to the Lognormal distribution on $\lambda$. The idea is to use a Gamma distribution to mimic the behavior of $h(\lambda;S)$ in Equation \eqref{compound}, also looking at the tail behavior of both distributions.

Both the Lognormal and the Gamma distribution are in fact skewed distributions, defined on the positive semi-axis, and characterized by a peculiar property: their coefficient of variation CV (the ratio of the standard deviation and the mean) is constant, and does not depend on both the mean and the standard deviation. In a $\text{Lognormal}(0,S^2)$ the CV is equal to $\sqrt{e^{S^2}-1}$, while for a positive random variable $Y$ following a $\text{Gamma}(\alpha,\beta)$ distribution with density
$$
\frac{\beta^\alpha}{\Gamma(\alpha)}y^{\alpha-1}e^{-\beta y}, \qquad \alpha>0, \beta>0,
$$
the CV is simply $\frac{1}{\sqrt{\alpha}}$.

From the point of view of extreme value statistics, both the Gamma and the Lognormal are heavy-tailed distributions, meaning that their right tail goes to zero slower than an exponential function, but not "true" fat-tailed, i.e. their tail decreases faster than a power law \cite{Rolski}. From the point of view of extreme value theory, both distributions are in the maximum domain of attraction of the Gumbel case of the Generalized Extreme Value distribution \cite{deHaan, Embrechts}, and not of the Fr\'echet one, i.e. the proper fat-tailed case. As a consequence, the moments of these distributions will always be finite.

\begin{figure}[htbp]
\begin{center}
\includegraphics[width=\linewidth]{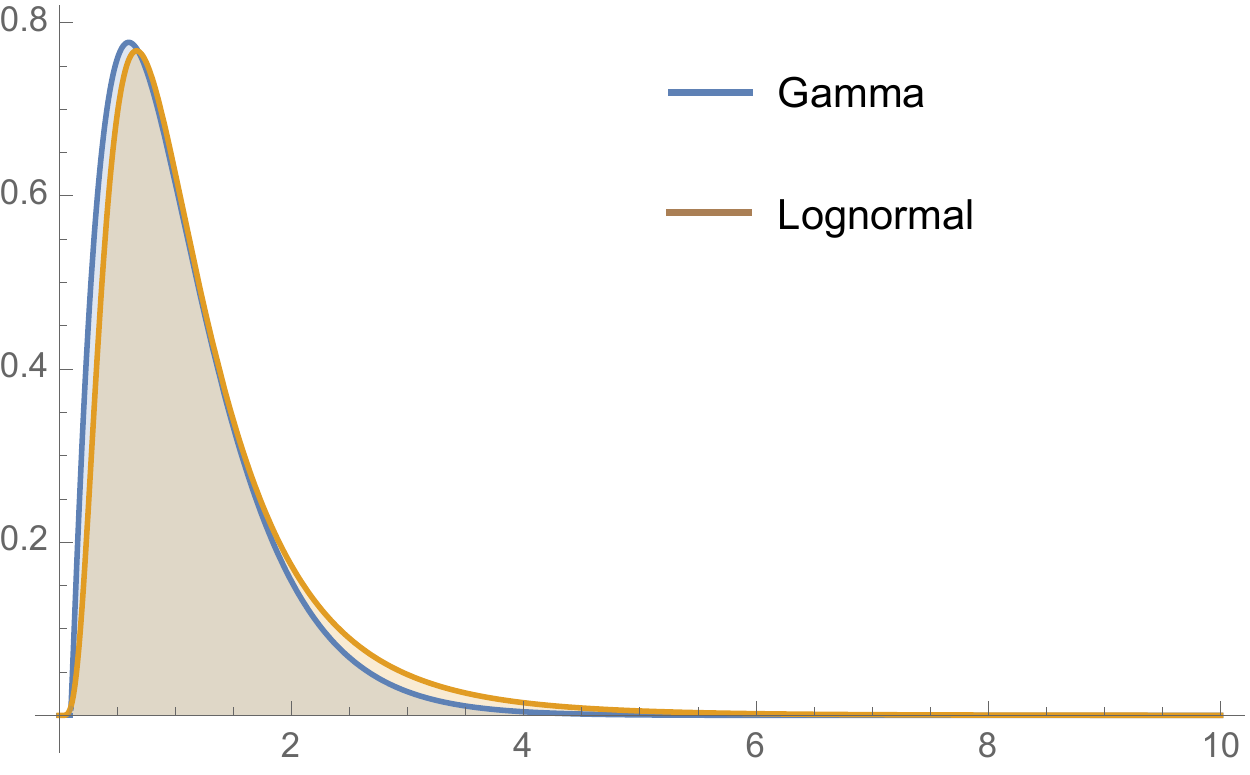}
\caption{Comparison of the density functions of a Gamma and a Lognormal sharing the same coefficient of variation of 0.7.}
\label{Comparison}
\end{center}
\end{figure}

As applied statisticians know \cite{Johnson}, from a qualitative point of view, it is rather difficult to distinguish between a Lognormal and a Gamma sharing the same coefficient of variation, when fitting data. In generalized linear models, it is nothing but a personal choice to use a Lognormal rather than a Gamma regression, more or less like choosing between a Logit and a Probit \cite{McCullagh}. In their bulk, a Lognormal and a Gamma with the same mean and standard deviation (hence the same CV) actually approximate quite well one another, as also shown in Figure \ref{Comparison}. The Gamma appears to give a little more mass to the smaller values, but the approximation is definitely good.

Interestingly, the Lognormal and the Gamma are also linked through the operation of exponentiation \cite{Consul}.

The main difference, when comparing a Lognormal and a Gamma sharing the same coefficient of variation is relative to the right tail. The Lognormal, in fact, shows a slightly heavier tail, whose decrease is slower, as also evident from Figure \ref{Comparison}. To verify analytically that the Lognormal tail dominates the one of the Gamma, we can have a look at their asymptotic failure rates \cite{Klugman}.

For a distribution $H(x)$ with density $h(x)$ and survival function $\bar{H}(x)$, the failure rate is defined as 
$$
r(x)=\frac{h(x)}{\bar{H}(x)}.
$$
The quantity $\tau=\lim_{x\to \infty} r(x)$ is called asymptotic failure rate. A distribution with a lower $\tau$ will have a thicker tail with respect to one with a larger asymptotic failure rate. When two distributions share the same $\tau$, conversely, more sophisticated analyses are needed to study their tail behavior \cite{Embrechts}. 

For a $\text{Gamma}(\alpha,\beta)$ it is easy to verify that 
$$r_{Gamma}(x)=\frac{e^{-\frac{x}{b}} \left(\frac{x}{b}\right)^{a-1}}{b \Gamma \left(a,\frac{x}{b}\right)},$$
while for a generic $\text{Lognormal}(\mu,\sigma^2)$ we have 
$$r_{Lognormal}(x)=\frac{\sqrt{\frac{2}{\pi }} e^{-\frac{(\mu -\log (x))^2}{2 \sigma ^2}}}{\sigma  x \text{erfc}\left(\frac{\log (x)-\mu }{\sqrt{2} \sigma }\right)},$$
where $\text{erfc}(\cdot)$ is once again the complementary error function. 

By taking the limits for $x\to \infty$, we see that $\tau_{Gamma}=\frac{1}{\beta}$, while $\tau_{Lognormal}=0$. Therefore, a Lognormal has a right tail which is heavier than that of a Gamma (actually all Gammas, given that $\tau_{Lognormal}$ does not depend on any parameter). A relevant consequence of this different tail behavior is that the Lognormal is bound to generate more extreme scenarios than the Gamma. This, combined with the fact that in the bulk the two distributions are rather similar--even if the Gamma slightly inflates the small values--allows us to say that we can use the Gamma as a lower bound for the Lognormal, when we center both distributions on the same coefficient of variation. 

Coming back to the CLT result of Section \ref{layers}, we can say that, in the limit, the precision parameter $\lambda$ can be taken to be approximately $\text{Gamma}(\alpha,\beta)$, where $\alpha$ and $\beta$ are chosen to obtain the same coefficient of variation of $\text{Lognormal}(0,S^2)$, that is $\alpha=\frac{1}{e^{S^2}-1}$ and $\beta=\frac{e^{-\frac{S^2}{2}}}{e^{S^2}-1}$.

In dealing with the precision parameter $\lambda$, moving from the Lognormal to the Gamma has a great advantage. A Normal distribution with known mean (for us $\mu=0$) and Gamma-distributed precision parameter has in fact an interesting closed form.

Let us come back to Equation \eqref{compound}, and let us re-write it by substituting the Lognormal density $h(\lambda;S)$ with an approximating $\text{Gamma}(\alpha,\beta)$ which we indicate with $h^\ast(\lambda;\alpha,\beta)$, obtaining
\begin{eqnarray} \label{compoundGamma}
g(x; \lambda,\alpha,\beta)&=&\int_{0}^\infty f(x; \lambda) h^\ast(\lambda;\alpha,\beta)\text{d} \lambda \\ 
&=& \frac{\beta^\alpha}{\Gamma(\alpha)\sqrt{2 \pi}} \int_{0}^\infty \lambda^{\alpha+\frac{1}{2}-1}e^{-\lambda\left(\beta+ \frac{1}{2}x^2 \right)} \text{d} \lambda. \nonumber
\end{eqnarray}
The integral above can now be solved explicitly, so that 
\begin{eqnarray} \label{compoundGamma2}
g(x; \alpha,\beta)&=& \frac{\beta^\alpha}{\Gamma(\alpha)\sqrt{2 \pi}}\frac{\Gamma \left( \alpha+\frac{1}{2}\right)}{\left(\beta+ \frac{1}{2}x^2 \right)^{\left( \alpha+\frac{1}{2}\right)}} \\
&=& \frac{\Gamma \left( \alpha+\frac{1}{2}\right)}{\Gamma \left( \alpha \right)\sqrt{2 b \pi}} \left(1+\frac{x^2}{2\beta}\right)^{-\left( \alpha+\frac{1}{2}\right)}. \nonumber
\end{eqnarray}
In Equation \eqref{compoundGamma2} we can recognize the density function of a non-standardized $t-$Student distribution with $2\alpha$ degrees of freedom, zero location and scale parameter $\beta$. As observed above, to guarantee the Gamma approximation to the $\text{Lognormal}(0,S^2)$, we set $\alpha=\frac{1}{e^{S^2}-1}$ and $\beta=\frac{e^{-\frac{S^2}{2}}}{e^{S^2}-1}$, where $S^2$ is the sum of the variances of the epistemic random errors. 

Interestingly, the t-Student distribution of Equation \eqref{compoundGamma2} is fat-tailed on both sides \cite{deHaan,Embrechts}, especially for small values of $\alpha$. Given that $\alpha$ decreases in $S^2$, which is the sum of the variances of the epistemic errors, hence a measure of the overall uncertainty, the more doubts we have about the precision parameter $\lambda$, the more the resulting t-Student distribution is fat-tailed, thus increasing tail risk. The actual value of $\alpha$ is indeed bound to be rather small. This result is in line with the findings of Section \ref{constant}.

Therefore, starting from a simple Normal distribution, by considering layers of epistemic uncertainty, we have obtained a fat-tailed distribution with the same mean ($\mu=0$), but capable of generating more extreme scenarios, and its tail behavior is a direct consequence of imprecision and ignorance. Since we have used the Gamma distribution as a lower bound for the Lognormal, we can expect that, with a Lognormal $\lambda$ the tails of $X$ will still be heavy and very far from normality.

\section{Discussion }

In Section \ref{layers} we started from a normally distributed random variable $X$ and we derived the effects of layering uncertainty on the standard deviation of $X$. We have analyzed different scenarios, all generating tails for $X$ that are thicker than those of the normal distribution we started from. 

Epistemic uncertainty was represented in terms of multiplicative errors, which can also be analyzed with a CLT argument leading to a Lognormal distribution for the precision parameter $\lambda=\frac{1}{\sigma^2}$. Given the impossibility of obtaining closed-form results for the Lognormal case, we used a Gamma approximation to obtain fat tails analytically, after noticing that the Lognormal will possibly generate even more extreme results, given that its tail dominates the one of the Gamma.

Now, the question is: how much do our results depend on the Normal-Lognormal-Gamma construction? 

 Centrally, the choice of the Normal distribution as starting point is not relevant. What really counts is the Lognormal emerging from the different layers of uncertainty on the parameter of choice, and the fact that such a Lognormal is riskier than a Gamma with the same coefficient of variation. In fact, if we start from an Exponential distribution with intensity parameter $\nu$, and on that $\nu$ we apply the same reasoning we developed on $\lambda$, then we will generate fat tails. In fact, the compounding of an Exponential distribution and a Gamma--which we use as ``lower bound" for the Lognormal--generates a Lomax, or Pareto II distribution, a well-known example of fat-tailed distribution \cite{Johnson}. If, conversely, we start from a Gamma we obtain a compound Gamma or a beta prime distribution (depending on parametrizations), other two cases of (possibly) fat-tailed distributions \cite{Kleiber}. Finally, even when dealing with discrete distributions, our approach may easily generate extremes (it is not correct to speak about fat tails with discrete distributions, given that some converge results used in the continuous case do not hold \cite{Embrechts}). For example, if we start from a Poisson with intensity $\mu$ and we apply our layers of uncertainty, what we get is a Negative Binomial (or something that resembles a Negative Binomial without the Gamma approximation), a skewed distribution, possibly with high kurtosis, used in risk management to model credit risk losses \cite{McNeil}.

\section{Applications and Consequences}

 Consequences in terms of risk management are clear: ignoring errors on errors induce a significant underestimation of tail risk. Those with forecasting less so: it is hard to conceive that even if past data shows thin-tailed properties, future data needs to be necessarily higher.
 
 We can also see how the out-of-sample can show degradation compared to in-sample properties: the future is necessarily to be treated as more fat-tailed than the past.
 
 More philosophically, our approach can help explain the central point in \textit{The Black Swan} \cite{Taleb}: one must deal with the future as if it were to deliver more frequent (or more impacting) tail events than what is gathered from our knowledge of the past\footnote{Taking all estimates and risks as an objective truth (i.e., attaching certitude to the estimate) is the greatest mistake a decision maker can make \cite{Rescher}. Uncertainty does not only emerge from the limits of our knowledge and the natural unpredictability (at least to a large extent) of complex systems and the future \cite{Knight, Gigerenzer, Shackle}, but it also permeates all our quantitative models for the real world, through our unavoidable errors. Understanding this limit is an essential step for effective risk management, as it help in shaping that sense of humility and precaution that make us prefer doubts and imprecisions to false certainties. }.
 
  %There are of course hindsight effect characterizing ``black swans" \cite{Taleb} comes from the late epiphany of having underestimated errors on errors and model risk, over-relying on estimates and toy models that make extremes and rare events impossible, without ever questioning the way in which the numbers we build theories and castles upon are obtained. 

\end{document}